\documentclass[singlespacing]{elsart}

\usepackage{natbib}
\usepackage{amssymb}
\usepackage{graphicx,amssymb}
\usepackage[version=3]{mhchem}

\begin{document}

\begin{frontmatter}

\title{Modeling a simple enzyme reaction with delay and discretization}
\author[ele,pff]{Jos\'e M. Albornoz\corauthref{cor}},
\corauth[cor]{Corresponding author.}
\ead{albornoz@ula.ve}
\author[pff]{Antonio Parravano}
\ead{parravan@ula.ve}

\address[ele]{Departamento de Electr\'onica y Comunicaciones, Facultad de Ingenier\'ia,
Universidad de Los Andes, La~Hechicera, M\'erida, M\'erida~5251, Venezuela.\\
Phone/Fax: +58-274-240 2907}

\address[pff]{Centro de F\'{\i}sica Fundamental,
Facultad de Ciencias, Universidad de Los Andes, \\ Apartado Postal 26
La~Hechicera, M\'erida, M\'erida~5251, Venezuela.\\
Phone/Fax: +58-274-240 1331}

\begin{abstract}

\begin{flushleft}
A comparison is made between conventional Michaelis-Menten kinetics and two models that take into account the duration of the conformational changes that take place at the molecular level during the catalytic cycle of a monomer. The models consider the time that elapses from the moment an enzyme-substrate complex forms until the moment a product molecule is released, as well as the recovery time needed to reset the conformational change that took place. In the first model the dynamics is described by a set of delayed differential equations, instead of the ordinary differential equations associated to Michaelis-Menten kinetics. In the second model the delay, the discretization inherent to enzyme reactions and the stochastic binding of substrates to enzimes at the molecular level is considered. All three models agree at equilibrium, as expected; however, out-of-equilibrium dynamics can differ substantially. In particular, both delayed models show oscillations at low values of the Michaelis constant which are not reproduced by the Michaelis-Menten model. Additionally, in certain cases, the dynamics shown by the continuous delayed model differs from the dynamics of the discrete delayed model when some reactant become scarce.
\end{flushleft}

\end{abstract}


\end{frontmatter}

\begin{flushleft}

\section{Introduction.}
For over a century, the conventional way to represent many enzyme reactions 
has been based on the Michaelis-Menten kinetics \citep{mm} (hereafter MM model), in 
which substrate S binds reversibly with enzyme E to form an enzyme-substrate 
complex ES, which is later transformed into product P and free enzyme E. Derivation of the Michaelis-Menten equation is based on the following scheme \citep{segel}:
\begin{center}
\ce{E + S <=>[\cmath{k_1}][\cmath{k_{-1}}] ES ->[\cmath{k_p}] E + P}
\end{center}

However, there is strong evidence suggesting that enzime activity depends on conformational changes in the enzyme structure \citep{bennett,hersch,haka,agmon,eisen,benko}. It is therefore reasonable to assume that a conformational change takes place from the moment an ES complex forms to the moment a product P is released; such change requires a certain time for its completion. This conformational change must then be reset before the enzyme is ready to bind substrate again, and so a recovery time must elapse once P is released. Single-enzyme studies \citep{lu,xie,eisen2,walter} and nuclear magnetic resonance spectroscopy studies \citep{eisen2,huang} support these notions. A scheme that considers enzyme regeneration is presented by \citet{english},
\begin{center}
\ce{E + S <=>[\cmath{k_1}][\cmath{k_{-1}}] ES ->[\cmath{k_2}] E^0 + P},\\ \ce{E^0 ->[\cmath{k_3}] E}
\end{center}
where $\mathrm{E^0}$ represents recovering enzyme. Conventional MM kinetics do not account in an explicit way for the times required by these conformational changes.

The aim of this paper is to present two models of enzyme catalysis, one continuous and one discrete, that take into account the times needed by enzymes to process ES complexes and to reset to its original state once products has been released; these models can be considered as an extension of conventional MM kinetics which includes delays. The main motivation behind the continuous delayed model is to establish the limitations of the MM description in non-equilibrium conditions. In order to examine the limitations associated to the continuous nature of the first model, we consider a discrete delayed model in which the catalytic cicle of each enzyme in the system is represented by a recursive function, hereafter referred to as the enzyme map. Previous work by \citet{stange1,stange2} present an analogous description of enzyme action in which each protein molecule is represented by a clock-like automaton that binds substrate S and transforms it into product P. The time needed by a single automaton to complete this transformation and return to its original waiting state represents the enzyme turnover time $\tau$. The conformational changes of the enzyme (and the different stages of the catalytic cycle) are represented by motion along an internal `phase' coordinate. Under the aproppriate conditions this model exhibits complex behavior such as oscillations and cluster formation; thus another motivation is to explore the emergence of such behavior in our model. Unlike other microscopic models based on stochastic simulation \citep{hasel,kierzek,rao,pucha}, our discrete model is completely deterministic, and each enzyme is represented by a map instead of a clock-like unit. 

In Section 2 the continuous delayed model is presented by means of a simple reaction, and a comparison is made between numerical integration provided by MM kinetics and those produced by the delayed model. Section 3 presents the equivalent discrete model; the results of the discrete simulation are compared with results obtained through the continuous delayed model. In the conclusions presented in Section 4 we highlight the respective merits and limitations of each model.

\section{MM vs continuous delayed model}
To illustrate the effect of delays on the dynamics of enzyme catalysis we consider the simple reaction \ce{A <=>[\alpha][\beta] B}, where conversion between two substrates $\mathrm{A}$ and $\mathrm{B}$ is catalyzed by two unidirectional monomeric enzymes $\alpha$ and $\beta$.

As mentioned in the introduction, each conformational change taking place along the catalytic cycle will require a certain time for its completion. We then distinguish three stages in the cycle: 1) the enzyme is free, waiting to bind substrate; 2) the enzyme has bound substrate forming an ES complex; 3) product P has been released and the enzyme is resetting to its initial conformation. The processing time $\tau_p$ required by stage 2 is a fraction $c$ of the turnover time $\tau$, while the recovery time $\tau_r$ associeted to stage 3 is $(1-c)\tau$. The turnover time is the time that elapses from the moment an ES complex forms until the moment the enzyme is ready to bind substrate again; such time is given by
\begin{equation}
\tau=\frac{\displaystyle 60}{\displaystyle V_{max}\mu_e\times 10^{-3}} \mbox{  [sec]}
\end{equation}
where $V_{max}$ is the maximum velocity of the enzyme in $\mu$mol/(min-mg of protein) and $\mu_e$ its molecular weight in Daltons. 

When processing and recovery times are considered the reaction is described by a set of four coupled delay differential equations (DDEs):
\begin{eqnarray}
\frac{d[\mathrm{A}]}{dt}&=&\frac{1}{V_r}\left\lbrace g\!_{_{\beta}}[\mathrm{B}(t-\tau\!_{_{\beta p}})]M\!_{_{\beta 1}}(t-\tau\!_{_{\beta p}})- g\!_{_{\alpha}}[\mathrm{A}(t)]M\!_{_{\alpha 1}}(t)\right\rbrace + \: [\dot{\mathrm{A}}_{ext}(t)] \label{dde1}\\
\frac{d[\mathrm{B}]}{dt}&=&\frac{1}{V_r}\left\lbrace g\!_{_{\alpha}}[\mathrm{A}(t-\tau\!_{_{\alpha p}})]M\!_{_{\alpha 1}}(t-\tau\!_{_{\alpha p}}) - g\!_{_{\beta}}[\mathrm{B}(t)]M\!_{_{\beta 1}}(t)\right\rbrace \label{dde2}\\
\frac{dM\!_{_{\alpha1}}}{dt}&=&\mu_{\alpha}\times 10^{-3} g\!_{_{\alpha}}\left\lbrace [\mathrm{A}(t -
\tau\!_{_{\alpha}})]M\!_{_{\alpha 1}}(t - \tau\!_{_{\alpha}}) - [\mathrm{A}(t)]M\!_{_{\alpha 1}}(t))\right\rbrace \label{dde3}\\
\frac{dM\!_{_{\beta1}}}{dt}&=&\mu_{\beta}\times 10^{-3} g\!_{_{\beta}}\left\lbrace [\mathrm{B}(t - \tau\!_{_{\beta}})]M\!_{_{\beta 1}}
(t - \tau\!_{_{\beta}}) - [\mathrm{B}(t)]M\!_{_{\beta 1}}(t))\right\rbrace\label{dde4} 
\end{eqnarray}

where 
$[\mathrm{A}]$, $[\mathrm{B}]$ are substrate concentrations in $\mu$M;
$M\!_{_{\alpha 1}}$, $M\!_{_{\beta 1}}$ represent the amount of free enzyme in milligrams and $\mu_{\alpha}$, $\mu_{\beta}$ are the molecular weights of enzymes $\alpha$, $\beta$ in Daltons.
The delays $\tau\!_{_{\alpha}}$ and $\tau\!_{_{\beta}}$ are their respective turnover times, whereas $\tau\!_{_{\alpha p}}=c\!_{_{\alpha}}\tau\!_{_{\alpha}}$, $\tau\!_{_{\beta p}}=c\!_{_{\beta}}\tau\!_{_{\beta}}$ are their respective processing times. $V_r$ is the volume in which the reacion takes place, in liters. The rates $g\!_{_{\alpha}}$ and $g\!_{_{\beta}}$ are given by $g_{_{i}}=V_{max_{i}}/K_{i}$ ($i=\alpha,\beta$), where $K_{i}$ is the Michaelis constant in $\mu$M. $[\dot{\mathrm{A}}_{ext}(t)]$ represents the rate at which substrate A is added to the system. Equations for $M\!_{_{\alpha 2}}(t)$, $M\!_{_{\alpha 3}}(t)$,  $M\!_{_{\beta 2}}(t)$ and $M\!_{_{\beta 3}}(t)$ can be easily obtained from $[\mathrm{A}(t)]$, $[\mathrm{B}(t)]$, $M\!_{_{\alpha 1}}(t)$, $M\!_{_{\beta 1}}(t)$. For instance, the term $\mu_{\alpha}\times 10^{-3} g\!_{_{\alpha}}[\mathrm{A}(t)]M\!_{_{\alpha 1}}(t)$ is the rate at which $M\!_{_{\alpha 2}}(t)$ appears, while $\mu_{\alpha}\times 10^{-3} g\!_{_{\alpha}}[\mathrm{A}(t-\tau\!_{_{\alpha p}})]M\!_{_{\alpha 1}}(t-\tau\!_{_{\alpha p}})$ is the rate of dissapearance of $M\!_{_{\alpha 2}}(t)$.

Equations (\ref{dde1}-\ref{dde4}) reduce to conventional MM equations 
\begin{eqnarray}
\frac{d[\mathrm{A}]}{dt}&=&\frac{1}{V_r}\left\lbrace  V_{max_{\beta}}\frac{[\mathrm{B}(t)]}{K_{\beta}+[\mathrm{B}(t)]}M\!_{_{\beta}} -V_{max_{\alpha}}\frac{[\mathrm{A}(t)]}{K_{\alpha}+[\mathrm{A}(t)]}M\!_{_{\alpha}}\right\rbrace +  [\dot{\mathrm{A}}_{ext}](t) \label{mm1}\\
\frac{d[\mathrm{B}]}{dt}&=&-\frac{d[\mathrm{A}]}{dt} \label{mm2},
\end{eqnarray}
when $\tau=0$. In this case, the amounts of free enzymes $M\!_{_{\alpha 1}}$ and $M\!_{_{\beta 2}}$ are given by
\begin{eqnarray}
M\!_{_{\alpha 1}}(t) &=&
\frac{M\!_{_{\alpha}}}{1+[\mathrm{A}(t)]/K_{\alpha}}\label{na1}\\
M\!_{_{\beta 1}}(t) &=& \frac{M\!_{_{\beta}}}{1+[\mathrm{B}(t)]/K_{\beta}}\label{nb1},
\end{eqnarray}
where $M\!_{_{\alpha}}$ and $M\!_{_{\beta}}$ are the total amounts of enzymes $\alpha$ and $\beta$ in milligrams, respectively. 
Notice that unlike the delayed model (Eqs. \ref{dde1}-\ref{dde4}), the MM equations (Eqs. \ref{mm1}-\ref{mm2}) implicitly assume that substrates are instantaneously transformed into products and that the free enzyme amounts are instantaneously determined by substrate concentrations. On the other hand, in the delayed model the rate at which substrate A appears is proportional to the rate at which substrate B was bound by enzyme $\beta$ at time $t-\tau\!_{_{\beta p}}$. Similarly, the rate at which free enzyme (e.g. $M\!_{_{\beta 1}}(t)$) appears at time $t$ equals the rate at which free enzyme dissappeared  a turnover time $\tau$ earlier (e.g. $M\!_{_{\beta 1}}(t-\tau)$).

When stationary conditions are reached in the continuous delayed model, the amounts of enzyme (in milligrams) in each stage are 

\begin{eqnarray}
\overline{M\!_{_{\alpha 1}}} &=& \frac{M\!_{_{\alpha}}}{1+\mu_{\alpha}\times 10^{-3}g\!_{_{\alpha}}\overline{[\mathrm{A}]}\tau\!_{_{\alpha}}}\\
\label{mest1}
\overline{M\!_{_{\alpha 2}}} &=& \frac{\mu_{\alpha}\times 10^{-3}g\!_{_{\alpha}}\overline{[\mathrm{A}]}M\!_{_{\alpha}}\tau\!_{_{\alpha p}}}{1+\mu_{\alpha}\times 10^{-3}g\!_{_{\alpha}}\overline{[\mathrm{A}]}\tau\!_{_{\alpha}}}\\
\overline{M\!_{_{\alpha 3}}} &=& \frac{\mu_{\alpha}\times 10^{-3}g\!_{_{\alpha}}\overline{[\mathrm{A}]}M\!_{_{\alpha}}\tau\!_{_{\alpha}}(1-c_{_{\alpha}})}{1+\mu_{\alpha}\times 10^{-3}g\!_{_{\alpha}}\overline{[\mathrm{A}]}\tau\!_{_{\alpha}}}
\end{eqnarray}
where the subscripts 1, 2, 3 correspond to free enzyme, occupied enzyme and recovering enzyme, respectively, and $\overline{[\mathrm{A}]}$ is the stationary value of concentration $[\mathrm{A}(t)]$. The same expressions can be used to determine $\overline{M\!_{_{\beta 1}}}$, $\overline{M\!_{_{\beta 2}}}$ and $\overline{M\!_{_{\beta 3}}}$ by replacing $\overline{[\mathrm{A}]}$ for $\overline{[\mathrm{B}]}$ and the parameters $M\!_{_{\alpha}}$, $\tau\!_{_{\alpha}}$, $\mu_{\alpha}$, $g\!_{_{\alpha}}$ and $c_{_{\alpha}}$ for those corresponding to enzyme $\beta$.

In the continuous delayed model the stationary substrate concentration $\overline{[\mathrm{A}]}$ corresponds to one of the solutions of a third degree polynomial when considering Eqs. (\ref{dde1}-\ref{dde4}) in stationary conditions. A similar equation must be solved to find the stationary concentration $\overline{[\mathrm{B}]}$. From the solutions of these polynomials, the stationary solution is the one satisfaying the restrictions $0<\overline{[\mathrm{A}]}<[\mathrm{S}_0]$, $0<\overline{[\mathrm{B}]}<[\mathrm{S}_0]$, where $[\mathrm{S}_0]$ is the total substrate concentration in the system. Note that at any time $[\mathrm{S}_0] =  [\mathrm{A}](t) + [\mathrm{B}](t) + [\alpha_2](t) + [\beta_2](t)$, where  $[\alpha_2](t)$ and $[\beta_2](t)$ are the concentrations of processing enzymes.

The corresponding stationary values of the substrate concentrations $\overline{[\mathrm{A}]}_{MM}$, $\overline{[\mathrm{B}]}_{MM}$ for the MM model (Eqs. (\ref{mm1}) and (\ref{mm2}))
are given by the solutions of two second-degree polynomials that satisfy $0<\overline{[\mathrm{A}]}<[\mathrm{S}_0]$, $0<\overline{[\mathrm{B}]}<[\mathrm{S}_0]$. Note that at any time $[\mathrm{S}_0] = [\mathrm{A}](t) + [\mathrm{B}](t)$. Since the MM description does not consider bound substrate, the continuous delayed model and the MM model agree in the stationary state when  $\overline{[\mathrm{A}]}_{MM}$ is compared to $\overline{[\mathrm{A}]}+\overline{[\alpha_2]}$; a similar relation applies to substrate B.

In order to establish a comparison between the delayed model and MM kinetics, numerical integration of Eqs. (\ref{dde1}-\ref{dde4}) and (\ref{mm1}-\ref{mm2}) was performed. Initial substrate concentrations are set to zero and all enzymes are initialized in their idle state. Later, substrate A is quickly added until a concentration $[\mathrm{S}_0]$ is reached according to $[\dot{\mathrm{A}}_{ext}(t)]= [\dot{\mathrm{A}}_{ext,max}] \exp(-(t-t_0)^2/\sigma)$, where $[\dot{\mathrm{A}}_{ext,max}]$ is the maximum rate of substrate addition reached at time $t=t_0$; the parameter $\sigma$ allows to control the time span during which significant amounts of substrate are added. Hereafter we adopt $\sigma= 0.1 \tau_{\alpha}^2$, and $t_0=2 \tau_{\alpha}$. Unless otherwise stated, the results shown correspond to the following parameter values: the reaction volume $V_r$ is set to 10.3 attoliters, corresponding to a \textit{T. brucei} glycosome \citep{orto}. For the sake of simplicity the reaction is made symmetric, i.e. $V_{max}=V_{max_{\alpha}} = V_{max_{\beta}} = 500$ $\mu$m/(min-mg of protein); the amount of enzyme is chosen to be $M\!_{_{\alpha}}$ = $M\!_{_{\beta}}$ = 1.66 attograms, a figure equivalent to 200 molecules of an enzyme with a molecular weight of 5000 Daltons. This corresponds to an enzyme concentration $[\alpha]=[\beta]=32.24$ $\mu$M. For these parameter values $\tau\!_{_\alpha}=\tau\!_{_\beta}=0.024$ sec. In the following example $K_M = K_{\alpha} = K_{\beta}$; for the DDE case we use $c=c_{_\alpha}=c_{_\beta}=0.5$.

Figure 1a shows a comparison between the evolution of [A] in the MM and DDE models for $K_M = 100$ $\mu$M and $[\mathrm{S}_0]=100$ $\mu$M. When the stationary regime is attained, the concentration [A] in the DDE case is below that of the MM case since in the DDE model a portion of substrate is bound in the form of ES complex with concentration $[\alpha_2]$. Figure 1b shows results when $K_M$ is reduced to 0.1 $\mu$M. The continuous thin curve correspond to the evolution of $[A] + [\alpha_2]$. Several new features are now apparent: a) The difference between MM and DDE free substrate concentrations widens because in the DDE model the portion of substrate in the form of ES complex increases as $K_M$ is reduced. And b) oscillations are present in the DDE case as a result of the history dependence of the system evolution, an usual behavior of systems described by delay differential equations \citep{parravan}. Note that a longer characteristic time is needed to reach equilibrium when $K_M$ is reduced; this is due to the fact that a reduction of $K_M$ increases the binding rate of both substrates while the fraction of available free enzymes decreases, but due to the symmetric cyclic nature of the reaction, substrate A is replenished at a faster pace.

Figure 2 presents the same comparison shown in Fig. 1b, but with $c$ lowered to $0.02$. As stated above, the MM model assumes the products are released instantaneously after substrate is bound by the enzime. Therefore, as expected, for $\tau\!_{_{p}}=c\tau \ll \tau$ substrate concentrations in the DDE model oscillate around those in the MM model.

Figure 3 shows a comparison between the MM and DDE models of the fraction of free enzyme $f_{\alpha 1}(t)$ and the amount of substrate A normalized to the total subtrate concentration $[\mathrm{S}_0]$ for $K_M=0.1$ $\mu$M, $[\mathrm{S}_0]=100$ $\mu$M and $c=0.5$. As substrate A is being added to the system the fraction of free enzyme $f_{\alpha 1}(t)$ in the DDE model shows a delay with respect to the MM model. This delay is the result of an important difference between MM kinetics and the delayed model: in the former, the amount of free enzyme is determined instantaneously by the substrate concentrations through Eqs. \ref{na1} and \ref{nb1}; this differs from the time-delayed evolution described by Eqs. \ref{dde3} and \ref{dde4}.

\section{Continuous delayed model vs a discrete model}
\citet{hess} proposed a discrete enzyme model in which each enzyme molecule is represented by a clock-like automaton with a phase that starts advancing once a substrate molecule binds to it, initiating a catalytic cycle whose duration represents the turnover time $\tau$. At a certain moment along this cycle the automaton releases a product, returning to its initial phase after a recovery time has elapsed. Enzyme reactions are thus represented in terms of a automata network, in which each automaton is coupled to the others through binding and release of intermediate products. Instead of a clock-like automaton, the discrete model proposed here assumes that the phase $x$ of each single enzyme along the catalytic cycle is determined by the recursive function $f(x)$ (hereafter the enzyme map) shown in Fig. 4. The map is a deterministic procedure to evolve the phase of the enzimes from its phase state $x_t$ at discrete time $t$ to $x_{t+1}$ at discrete time $t+1$; that is $x_{t+1}=f(x_t)$.

The map has a chaotic region defined in $0\leq x\leq1$, which corresponds to the idle state of the enzyme (e.g. $\alpha_1$ for enzyme $\alpha$). Iterations start in this region from an initial random phase value $x_0$; the average number of iterations that transcur in the chaotic region is inversely proportional to $p$, the width between the two intersections of the map at $f=1$ ($p=0.25$ in Fig. 4). When an iterate $x_t$ in the chaotic region falls into the range $(1-p)/2<x_t<(1+p)/2$ the enzyme is assumed to bind a substrate molecule: the next iterate exits the chaotic region and enters the laminar region. The laminar region is divided in two sections: the section from $x=1$ to $x=1+cb$ corresponds to the processing of the substrate by the enzyme (i.e. $\alpha_2$) whereas the section from $x=1+cb$ to $x=1+b$ corresponds to the recovery process of the enzyme (i.e. $\alpha_3$). Once $x_t>1+cb$ a product molecule is released, as indicated in Fig. 4. The region of the map between $x=1+b$ and $x=1+b+a$ allows to reinject in a single step the iterate into the chaotic region, where the enzyme is ready to bind a substrate again. The enzyme map can also be used to model more complicated structures and functions such as oligomers and competitive/non-competitive inhibition; these are properties that will be presented elsewhere.

For a fixed value of $a$ (e.g. $a=1/10$), the two parameters of the map $b$ and $c$, and the variable $p$ can be adjusted to obtain an equivalence with the continuous delayed model. Note that the number of steps in the laminar region is $b/a$; then if each time step corresponds to a unit time $\Delta t$, the turnover time is $\tau=\Delta t\, b/a$. The time $\Delta t$ associated to each iteration must be small enough to produce a significant number of iterations in the laminar region; once a given $\Delta t$ is chosen the length of the laminar region is set according to $b=\tau\,a/\Delta t$. The parameter $c=\tau_p/\tau$ is the same one defined for the continuous delayed model. Variable $p$ is related to the average time $\tau_1$ that an enzyme remains in its idle state and is assumed proportional to the number $n_s$ of substrate molecules in the simulation volume (i.e. $\tau_1=\Delta t/p=1/(k_1 n_s)$, where $k_1$ is the \ce{E + S ->[k_1] ES} enzyme rate constant). The number of susbstrate molecules that corresponds to a given concentration [A] (in $\mu$M) is $n_A=V_r[\mathrm{A}]10^{-6}N_A$, where $N_A$ is Avogadro's number. When [A] = $K_M$ the velocity of the enzyme is halved; in this condition the time spent in the chaotic region must be equal to the time spent in the laminar region; that is $\tau_1=\tau$. Therefore, $p=\Delta t\, k_1\, n_s$ with $k_1=V_{max}\mu_e/(60V_rK_M 10^{-3}N_A)$. Finally, the number of enzymes $n_e$ that corresponds to a given mass of enzyme $m_e$ in mg is $n_e=m_eN_A/(\mu_e 10^3)$. 

By using an array of such maps it is possible to consider the effect of the discrete nature of reactants on the dynamics of the system and analize the conditions under which the continuous delayed model is no longer appropriate to describe very small systems.

Figure 5 shows integration results for the delayed continuous model and its equivalent discrete counterpart. In the DDE simulation we use the same parameter values as in Figs. 1 and 2, but with $K_M=K_\alpha=K_\beta=0.1612$ $\mu$M. In the system volume of 10.3 attoliters this concentration correspond to $1$ substrate molecule; on the other hand the total substrate concentration $[\mathrm{S}_0]$ = 100 $\mu$M corresponds to 620 substrate molecules. As before, there are 200 enzyme molecules of each type with a turnover time $\tau=0.024$ s and a molecular weight of 5000 Da. Therefore, in the discrete simulation we must follow the phase evolution of 400 maps, 200 for each type of enzime. If we adopt $\Delta t=1$ $\mu$s and $a=1/10$, the transit in the laminar region occurs in 24000 iterates (i.e. $b=2400$). As the reaction progresses the number of available free subtrate molecules ($n_A$ and $n_B$) changes, and so the value of $p$. For the maps representing enzime $\alpha$ we have $p_\alpha=n_A \times \Delta t\,V_{max}\mu_e/(60V_rK_\alpha 10^{-3}N_A) =(\Delta t/\tau)(n_A/n_\alpha) \simeq n_A/24000$, where $n_\alpha$ is the number of substrate molecules corresponding to concentration $K_\alpha$. A similar relation holds for $p_\beta$. When the $n_A/n_\alpha$ ratio is 1 the number of iterations spent in the chaotic region equals the number of iterations spent in the laminar region, so the enzyme will operate at half its maximum velocity.

As can be seen in Fig. 5, agreement between both models is excellent, but there is one caveat: care must be taken so that the time step $\Delta t$ is small enough to guarantee that always $p<1$; otherwise the results provided by the discrete model might disagree from the results of the continuous delayed model. In other words, $\Delta t$ must be less than $\tau\,n_\alpha/n_S$, where $n_S$ is the number of substrate molecules corresponding to the total substrate concentration $[\mathrm{S}_0]$.

Unlike the discrete model, the DDE approach is expected to fail when the number of substrate molecules in the reaction volume is small. As an example of this limitation, Fig. 6 shows the evolution of free substrate $A$ in both models when $[\mathrm{S}_0]$ is reduced to 32.24 $\mu$M, a concentration equivalent to 200 substrate molecules, half the number of enzyme molecules. The evolution of [A] shows how the results provided by the discrete model tend to follow those corresponding to the delayed continuous model; however fluctuations in [A] can be seen as substrate molecules are bound and released by the enzymes. Moreover, synchronization between the delayed continuous model and the discrete model is lost as the reaction progresses. This is the type of situation that can be expected in conditions of substrate starvation or when the reaction volume is reduced while maintaining $[\mathrm{S}_0]$ and the amount of enzyme.

As previoulsy mentioned, in the discrete model it is possible to follow the phase evolution of every single enzyme in the system. We show this by examining the same simple reaction used so far but with parameter values for which sustained oscillations occur; they are $V_r=5.15$ attoliters, $V_{max_{\alpha}}=500$ $\mu$m/(min-mg), $V_{max_{\beta}}=1000$ $\mu$m/(min-mg), $c\!_{_{\alpha}}=0.5$, $c\!_{_{\beta}}=0.02$, $K_{\alpha} = 1$ $\mu$M, $K_{\beta} = 0.5$ $\mu$M, and $[\mathrm{S}_0]=150$ $\mu$M (equivalent to 465 substrate molecules). The number and molecular weight of the enzymes remains the same as in previous cases. Figure 7 shows the phase histograms for enzymes $\alpha$ for two time moments separated by $\tau_\alpha/4$, illustrating how the phases of enzimes $\alpha$ tend to be synchronized and how the cluster phase advances through the catalytic cycle. In the first histogram there are some enzymes waiting to bind substrate, while in the second histogram these enzymes have already bound a substrate. 

\section{Conclusions}
We have included processing and recovery times in enzyme kinetic models in order to account in a simple way for the conformational changes that a monomer undergoes during the catalytic cycle and its effects on the resulting dynamics. In certain cases, the dynamics exhibited by the continuous delayed model differs drastically when compared to that of MM kinetics. Examples found in the literature indicate that oscillations in enzyme reactions require some strong form of non-linearity: inhibition-activation mechanisms, presence of feedback loops, interaction between the elements of a network, or cooperativity between enzyme subunits \citep{selkov,gold,laub,orto,chicka}. However, the emergence of oscillations in enzyme reactions can also be promoted by the delayed dynamics. 

Unlike the MM model, the delayed model accounts for that portion of substrate that is present in the system as ES complex; thus stationary substrate concentration values agree when the sum of free and bound substrate concentrations in the continuous delayed model are compared with free substrate concentrations in the MM model. Another important difference between MM kinetics and our model is that the fraction of free enzyme in the continuous delayed model follows a time-delayed evolution, while in the MM case such fraction is instantaneously determined by substrate concentrations.

The proposed discrete model agrees with the continuous delayed model when large enough number of substrate molecules are considered. When this number is small the discrete model shows the expected fluctuations that are associated to the discrete nature of reactions at microscopic scales. These effects can be important when considering reactions that take place in very small volumes such as those of organelles.

The discrete model offers the possibility of following the evolution of each single enzyme, and therefore allows for the study of situations in which self-organization phenomena arise, such as synchronization and clustering. 

\section*{Acknowledgements}
The authors would like to thank Dr. Juan Luis Concepci\'on and Dr. Luisana Avil\'an for their valuable opinions and insight during the many fruitful discussions we had.

\end{flushleft}

\eject

\begin{flushleft}

Fig. 1. a) [A] evolution for MM and DDE models when $V_{max}=V_{max_{\alpha}} = V_{max_{\beta}} = 500$ $\mu$m/(min-mg), $K_M = K_{\alpha} = K_{\beta}=100$ $\mu$M, $[\mathrm{S}_0]=100$ $\mu$M, $m_{\alpha}$ = $m_{\beta}$ = 1.66 attograms, $c=c_{\alpha}=c_{\beta}=0.5$ and $V_r=10.3$ attoliters. b) Same situation with $K_M = K_{\alpha} = K_{\beta}=0.1$ $\mu$M (see text).

Fig. 2. Same as in Fig. 1b but with $c=c_{\alpha}=c_{\beta}=0.02$.

Fig. 3. Fraction of free enzyme $\alpha$ and normalized concentration [A] for MM and DDE models for the parameter values used in Fig. 1b.

Fig. 4. The enzyme map $f(x)$ (see text).

Fig. 5. Evolution of the amount of free substrate A for DDE model and the discrete model. Left hand y-axis shows concentration values while right hand y-axis shows corresponding number of susbstrates. Paramater values are $V_{max}=V_{max_{\alpha}} = V_{max_{\beta}} = 500$ $\mu$m/(min-mg), $K_M = K_{\alpha} = K_{\beta}=0.1612$ $\mu$M, $[\mathrm{S}_0]=100$ $\mu$M, $m_{\alpha}$ = $m_{\beta}$ = 1.66 attograms, $c=c_{\alpha}=c_{\beta}=0.5$, $\Delta t = 1$ $\mu$s and $V_r=10.3$ attoliters.

Fig. 6. Same as Fig. 5 but with $[\mathrm{S}_0]=32.24$ $\mu$M.

Fig. 7. Phase histograms for enzyme $\alpha$. Parameter values are $V_{max_{\alpha}}=500$ $\mu$m(min-mg), $V_{max_{\beta}}=1000$ $\mu$m/(min-mg), $c_{\alpha}=0.5$, $c_{\beta}=0.02$, $K_{\alpha} = 1$ $\mu$M, $K_{\beta} = 0.5$ $\mu$M, and $[\mathrm{S}_0]=150$ $\mu$M and $V_r=5.15$ attoliters.

\end{flushleft}

\vfil

\eject

\begin{figure}
\includegraphics[width=15cm,angle=-90]{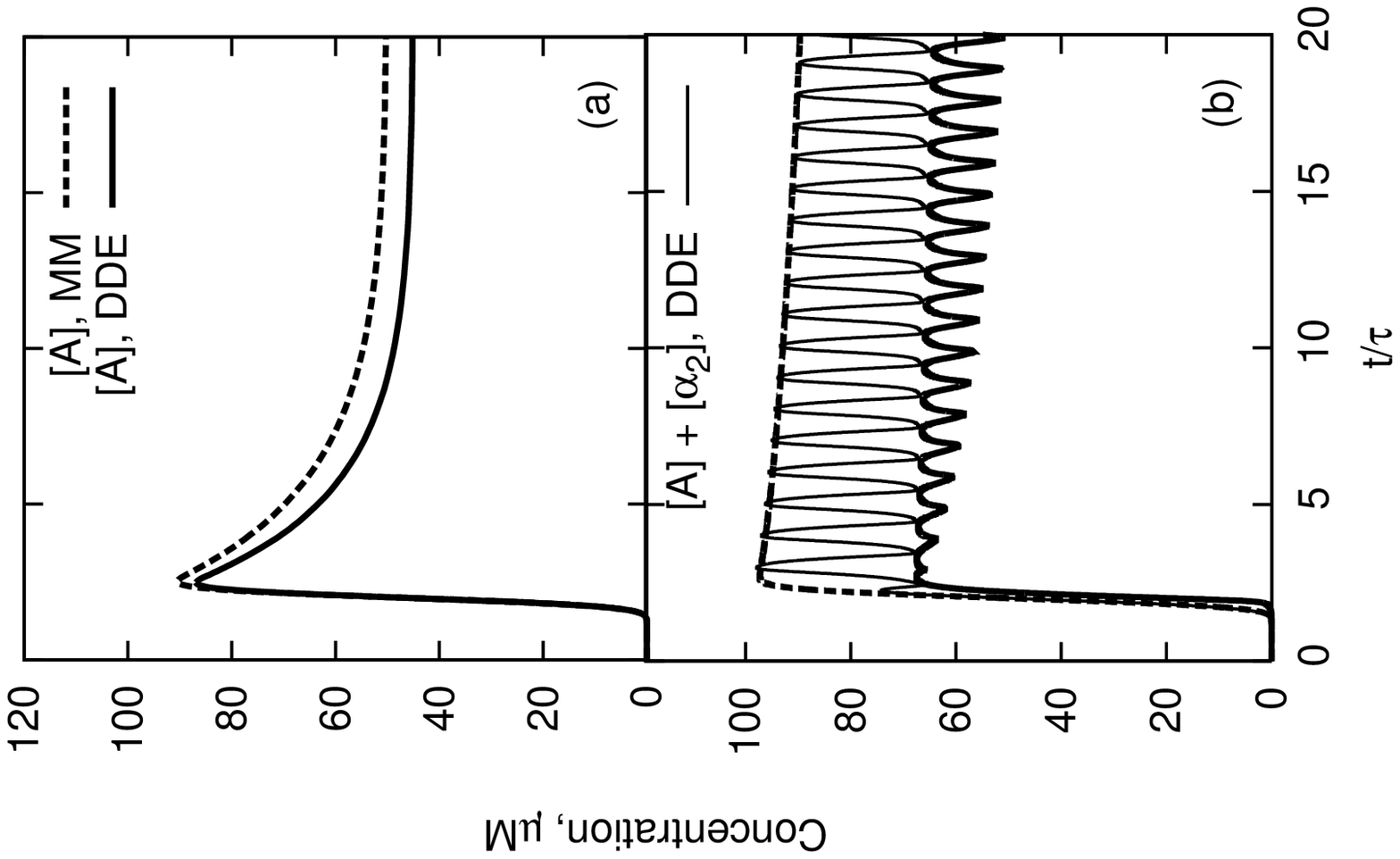}
\caption{}
\end{figure}

\begin{figure}
\includegraphics[width=11cm,angle=-90]{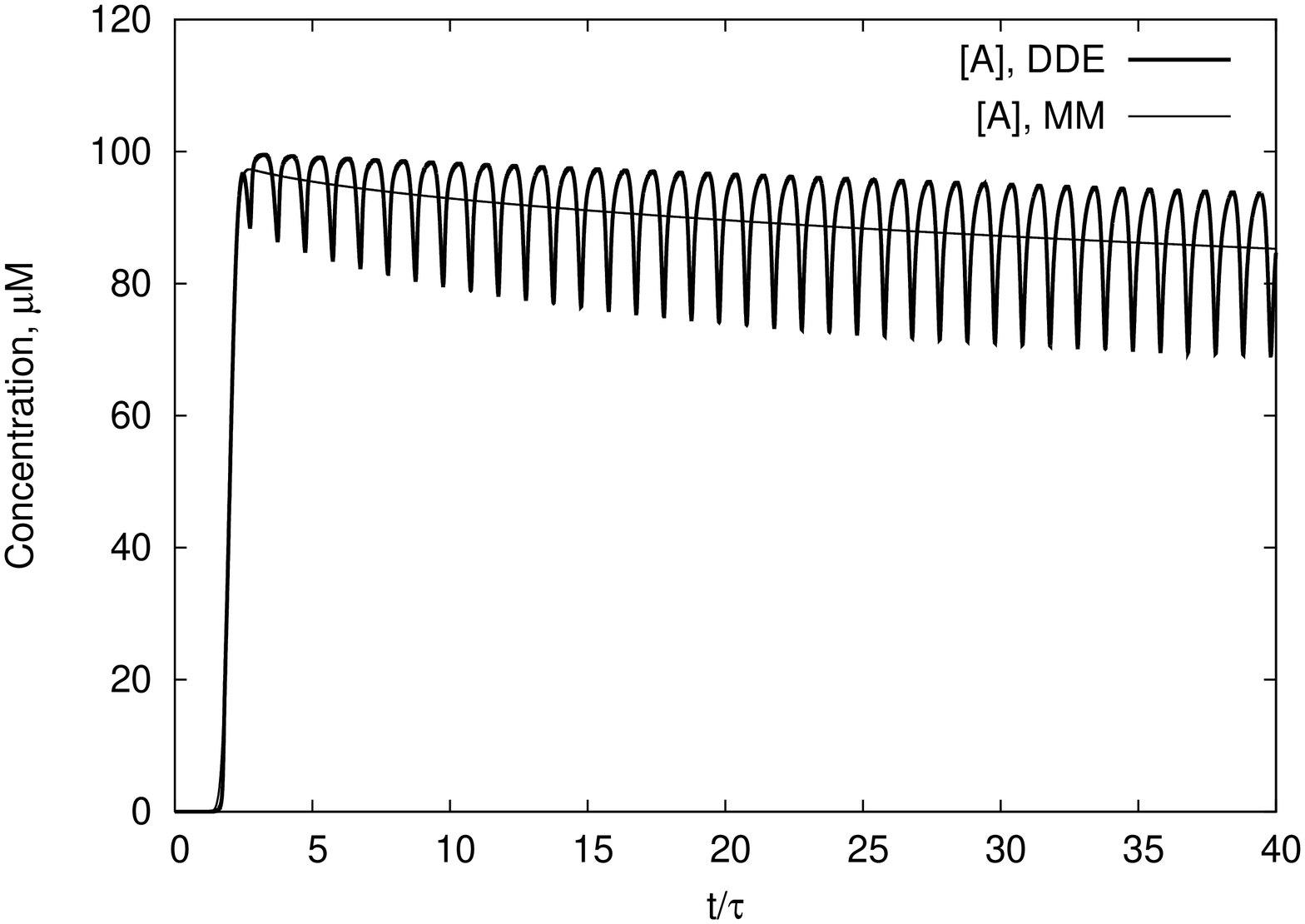}
\caption{}
\end{figure}

\begin{figure}
\includegraphics[width=11cm,angle=-90]{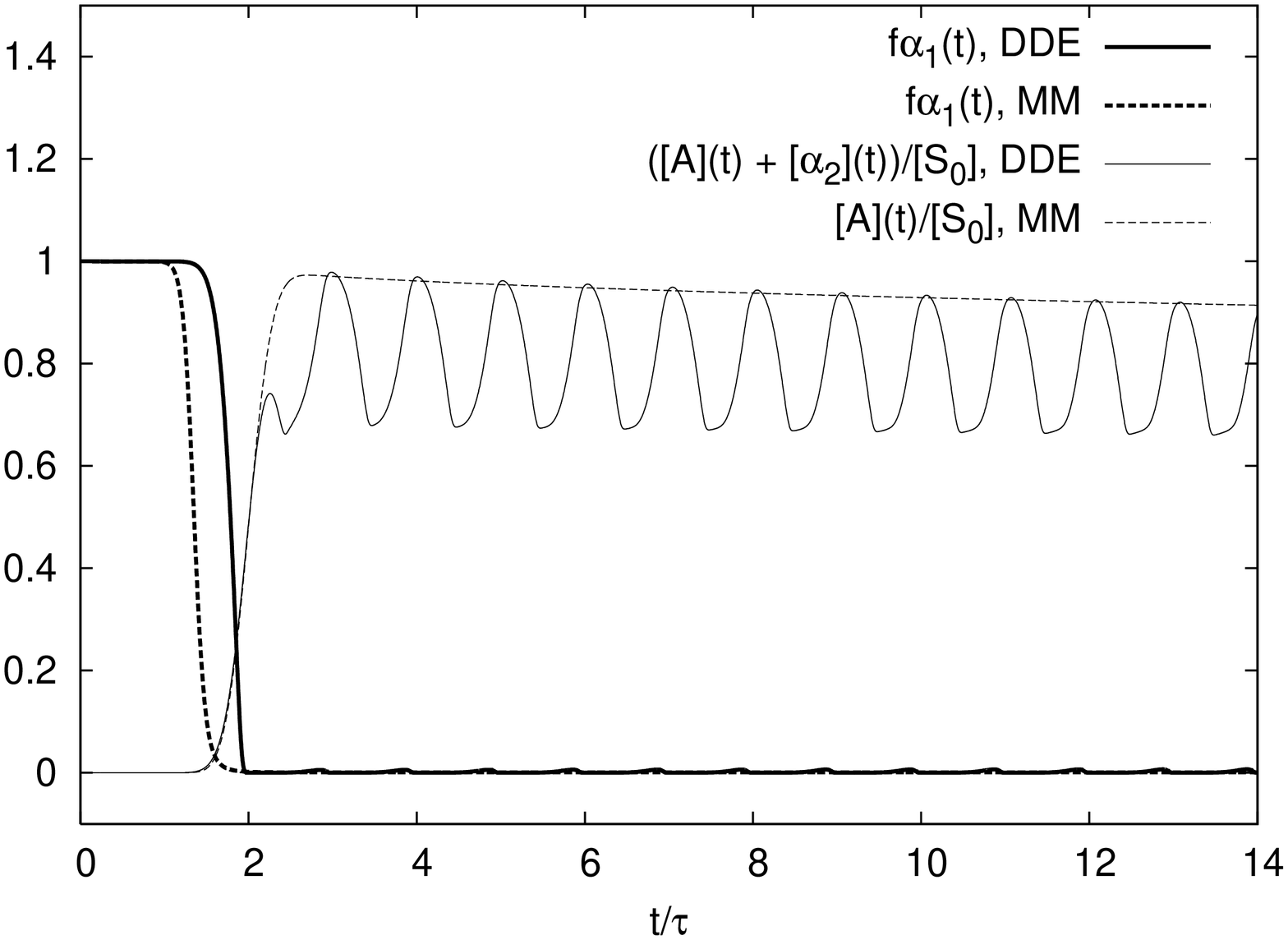}
\caption{}
\end{figure}

\begin{figure}
\includegraphics[width=11cm,angle=0]{fig4.eps}
\caption{}
\end{figure}

\begin{figure}
\includegraphics[width=11cm,angle=-90]{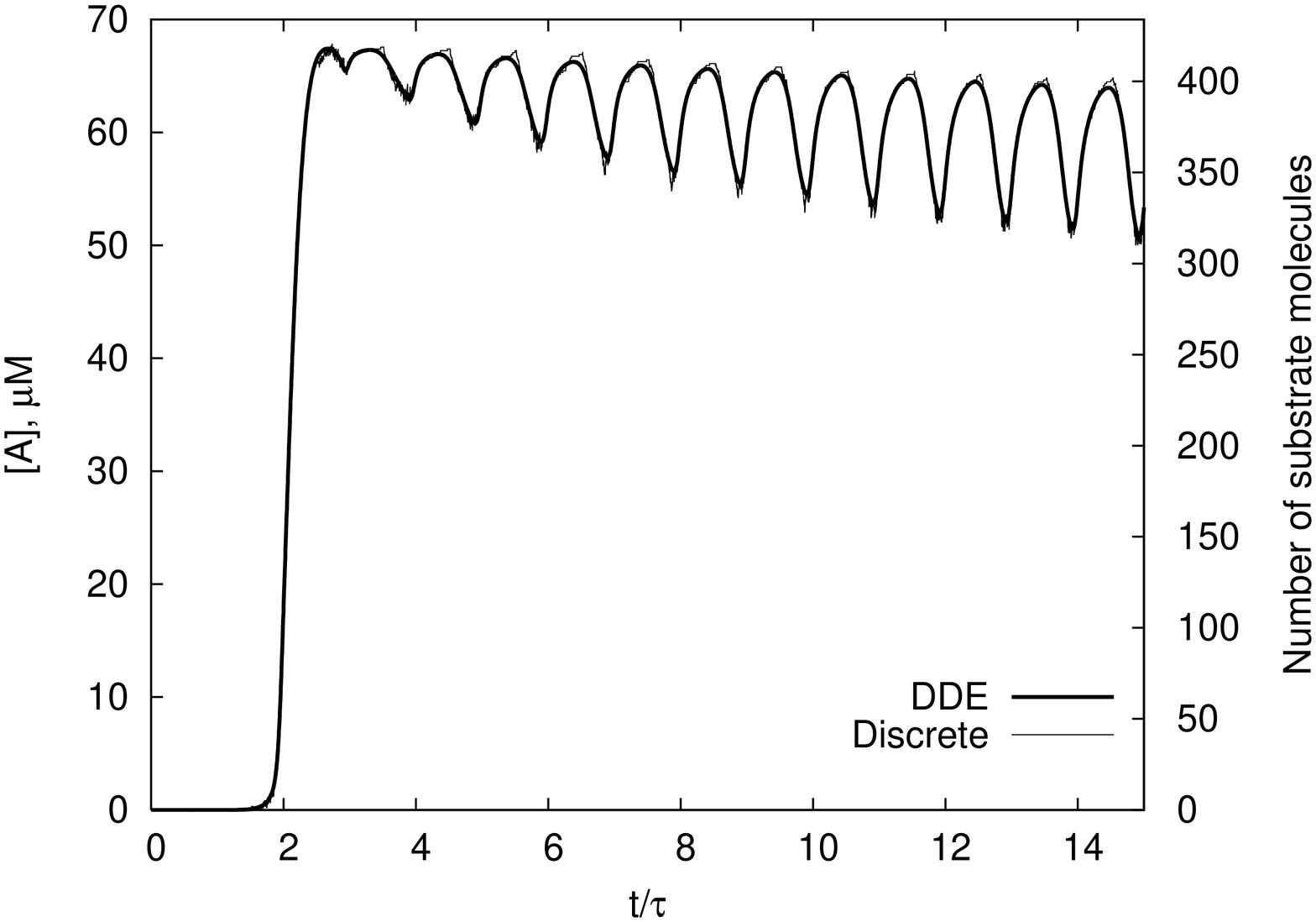}
\caption{}
\end{figure}

\begin{figure}
\includegraphics[width=11cm,angle=-90]{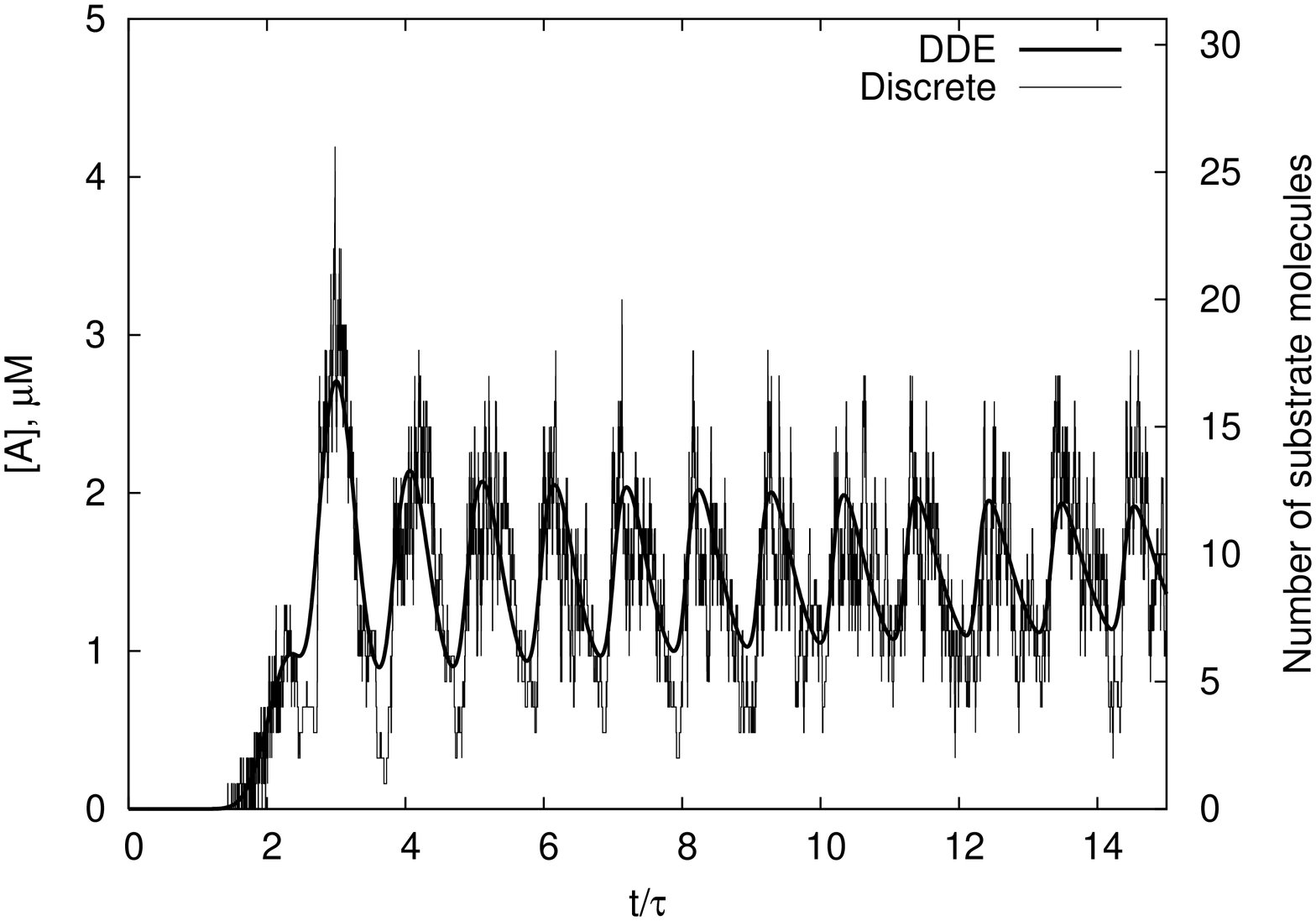}
\caption{}
\end{figure}

\begin{figure}
\includegraphics[width=15cm,angle=0]{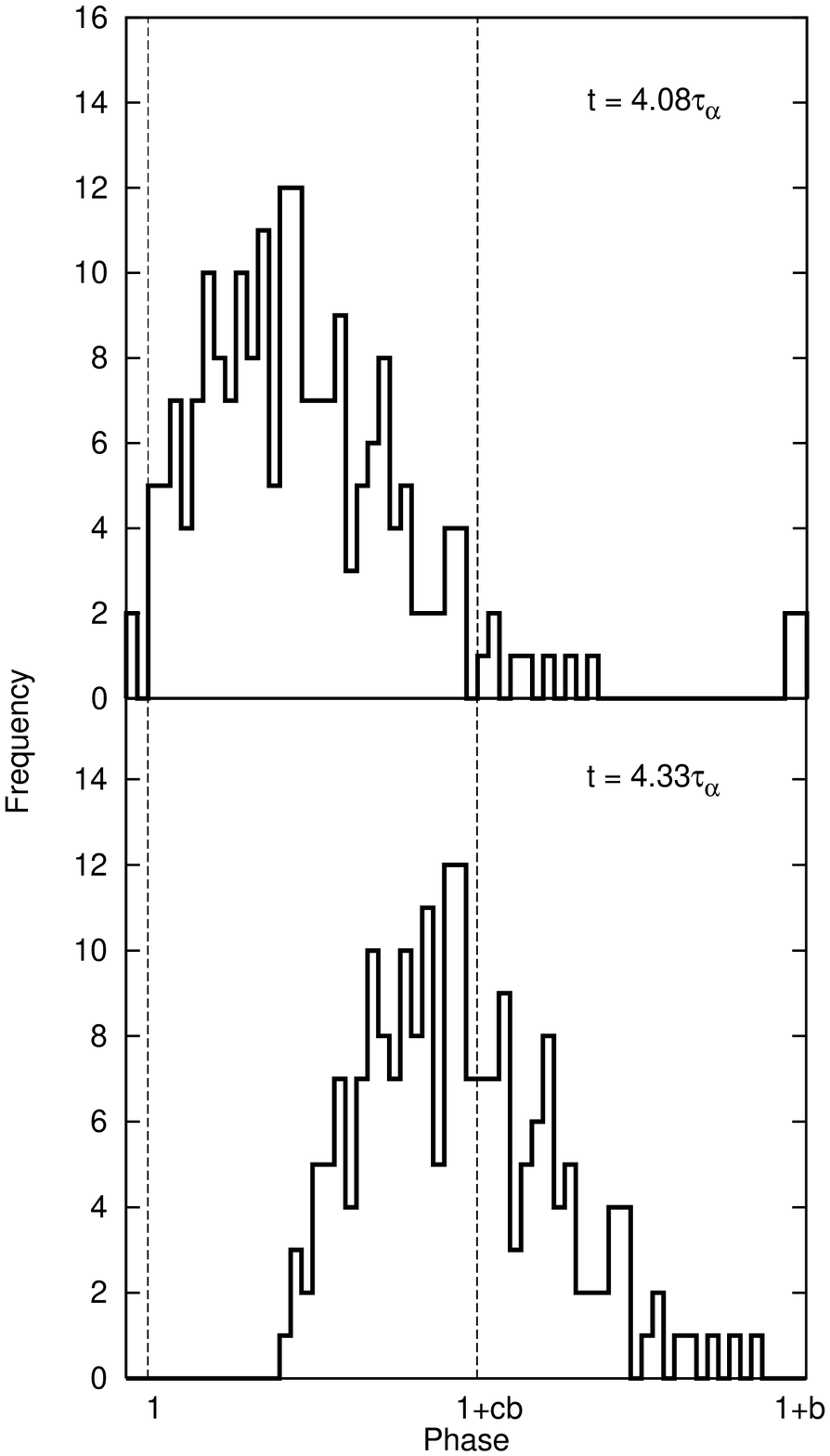}
\caption{}
\end{figure}

\end{document}